\documentclass[manuscript]{aastex}

\usepackage{graphics,color}
\usepackage{ulem}
\definecolor{red}{rgb}{1,0,0}

\shorttitle{Super-Fast-Rotator}
\shortauthors{Chang et al.}
\begin{document}
\title{A Quick Test on Rotation Period Clustering for the Small Members of the Koronis Family}

\author{Chan-Kao Chang\altaffilmark{1}; Hsing-Wen Lin\altaffilmark{1}; Wing-Huen Ip\altaffilmark{1,2};}

\altaffiltext{1}{Institute of Astronomy, National Central University, Jhongli,Taiwan}
\altaffiltext{2}{Space Science Institute, Macau University of Science and Technology, Macau}

\email{rex@astro.ncu.edu.tw}

\begin{abstract}
Rotation period clustering in prograde/retrograde rotators might indicate the preliminary indication of the
Slivan state in the Koronis family as a result of the YORP effect. We follow the general scenario of
dispersion in semimajor axis of the asteroid family members to separate prograde and retrograde
rotators in the Koronis family. From the available rotation periods obtained from PTF/iPTF, we can not found the rotation period clustering of objects with $H \gtrsim 12$ mag in the Koronis family. This could be
the result of the intermittent collisional process of small asteroids ($D \lesssim 20$ km) which leads to
astray Yarkovsky drifting. Measurement of the pole orientations of our sample will verify our preliminary result and validate our method.

\end{abstract}


\section{Introduction}
One of the interesting discoveries from the studies of asteroid rotation is the ``Slivan State''
\citep{Slivan2002}, which is the spin vector alignment and rotation period clustering of
relatively large asteroids ($H \lesssim 12$ mag and $D \gtrsim 20$ km) in the Koronis family, which is a relatively old family \citep[1.8--3 Gyr][]{Binzel1988, Greenberg1996, Spoto2015} at outer main-belt (semimajor axis of $\sim$ 2.87 AU).  Later, \citet{Vokrouhlicky2003} used the Yarkovsky-O'Keefe-Radzievskii-Paddack effect \citep[YORP;][]{Rubincam2000} to explain how the spin states of the Koronis family members could be driven within 2--3 Gyr into the $s_6$ spin-orbit secular
resonance between the precession rate of the asteroid's spin axis and Saturn's longitude of node. This discovery reveals the subtle influence of thermal effect on asteroid rotation. Such spin vector alignment
has also been reported in the Flora family \citep{Kryszczynska2012, Kryszczynska2013}. However, this
was not seen in the Maria family \citep{Kim2014}.

Determination of asteroid's spin vectors is extremely time-consuming as it requires intensive lightcurve observations at various phase angles. Therefore, it is difficult to verify the ``Slivan State'' in
asteroid families. However, as \citet{Slivan2002} pointed out, rotation period clustering should be
observed in the Koronis family once the ``Slivan State'' is established, in which the prograde
rotators are confined in rotation periods between 7--10 hr and the retrograde rotators are driven to rotation periods of $< 5$ or $> 13$ hr. From this point of view, rotation period clustering can be used as proxy of the ``Slivan State'' for the Koronis family even if the rotational directions of family members are determined.

It is believed that an asteroid family was formed in a catastrophic break-up of its parent body
which resulted in an initial dispersion of the fragments' semimajor axes. Further dispersion occurred as the family members drifted away from the family's center due to the Yarkovsky effect. According to
\citet{Vokrouhlicky1998, Vokrouhlicky1998b, Farinella1999, Vokrouhlicky1999}, the prograde/retrograde
rotators would increase/decrease their semimajor axes (hereafter, outer/inner sides). Therefore,
a simple assumption can be made on the family members outside the initial dispersion in semimajor axis.
Those at the inner side (hereafter, inner-side members) are expected to be mostly retrograde rotators and
those at the outer side (hereafter, outer-side members) should be mostly prograde rotators. In this way, we should be able to see more outer-side members of the Koronis family in rotation periods of 7--10 hr and correspondingly less
inner-side members in rotation periods of 5--13 hr.

The PTF/iPTF\footnote{http://ptf.caltech.edu/iptf} is an all-sky time-series survey, which has
six-year archived data and has been conducting several dedicated observations for asteroid spin-rate
study. The large number of asteroid rotation periods obtained from archived data and dedicated
observations of PTF/iPTF is very useful to test the aforementioned idea on the Koronis family
\citep{Chang2014, Chang2015, Waszczak2015}. In Section 2, we describe our algorithm in detail. The
data used in the analysis are given in Section 3. Results and discussion are presented in Section
4. A summary and conclusion can be found in Section 5.

\section{Method}\label{method}
An asteroid family is a group of asteroids, which share similar proper elements of semimajor axis
($a$), eccentricity ($e$) and inclination ($i$), resulting from a catastrophic break-up event. When an
asteroid family was formed, the fragments followed the size-velocity relation, which
describes the relationship between the sizes of asteroids and their velocity dispersions. Different velocity
dispersions in the break-up events would produce different initial dispersions in semimajor axis \citep[$\Delta a$; see][ and references therein]{Zappala2002}. This can be used in age determination of asteroid families \citep[see examples in][]{Vokrouhlicky2006}. To calculate $\Delta a$, we follow the approach of \citet{Cellino1999, Vokrouhlicky2006},
\begin{equation}
  \Delta a = \frac{2}{n} V_T + O(e),
\end{equation}
where $n$ is the mean motion of the break-up fragment, $V_T$ is its transverse velocity, and $O(e)$
refers to other terms related to eccentricity, which is neglected when circular orbit is considered.
Moreover, we assumed size-velocity relationship
\begin{equation}
  V_T = V_0(\frac{D_0}{D}),
\end{equation}
for the velocity dispersion of a fragment with size $D$, where $V_0$ is the velocity dispersion of our reference diameter $D_0 = 5$ km. Therefore,
\begin{equation}
  \Delta a \sim \frac{2}{n} V_0 (\frac{D_0}{D}).
\end{equation}
In addition, the diameter $D$ is related to $H$ by
\begin{equation}
  D = \frac{1329}{\sqrt{p_v}}10^{-0.2H},
\end{equation}
where $p_v$ is geometric albedo and we adopt the average value $p_v = 0.24$ for the Koronis family \citep{Masiero2011}.
We can quantify $\Delta a$ on the plot of $a$ vs. $H$. In addition, the orbital motion of the family members are modified
by the Yarkovsky effect, where prograde and retrograde rotators would move outward and inward along semimajor axis, respectively.
After certain amount of time, some members would leave $\Delta a$. Therefore, we can expect that family members
located beyond the $\Delta a$ interval should be subdivided in two categories: those moving to greater (i.e., outer-side members)
and smaller (i.e., inner-side members) semi-major axes are expected to be mostly prograde and retrograde rotators, respectively. Therefore, a clustering of
the rotation period around 7 - 10 hr is expected among the former objects in case they are in a Slivan state,
whereas the objects moving to smaller semi-major axis should exhibit a lack of rotation periods between 5 and 13 hr.

\section{Data}
\subsection{Family Member Selection}
\citet{Nesvorny2012} used analytic/synthetic proper elements to calculate asteroid dynamical family
memberships with the Hierarchical Clustering Method \citep[HCM;][]{Zappala1990, Zappala1994} and
generated the data set, Nesvorny HCM Asteroid Families V2.0. To select the Koronis family members, we
adopt both catalogs calculated with analytic and synthetic proper elements. However, we
exclude the members of the Karin family, a sub-family in the Koronis family, from the selected
sample. To ensure the membership of the aforementioned selected objects, we also use the asteroid family catalog of \citet{Milani2014} as a reference. In addition, the SDSS colors \citep{Parker2008} of these dynamically selected members, if available, are used to identify the spectral types of the family members.

\subsection{Asteroid Rotation Periods}
The rotation periods of the Koronis family members used in this study are collected from two data sets
and each is briefly described below.

\textit{The archived PTF asteroid rotation periods} \citep[][; hereafter PTF-Waszczak]{Waszczak2015}.
The data set used 5-yr archived data of PTF/iPTF\footnote{Intermediate Palomar Transient Factory; http://ptf.caltech.edu/iptf} to obtain 54,296 sparsely-sampled asteroid lightcurves, from which $\sim$ 8,300 reliable rotation periods were determined by fitting rotation and phase-function model simultaneously. The data set is sensitive to detect spin rate of $\lesssim 10$ rev/day (see Figure 7 in \citet{Waszczak2015}). Most objects in the data set are main-belt asteroids of $D \gtrsim 2$ km. Therefore, this data set includes plenty of members of asteroid families and is the main source used in this study to verify rotation period clustering of asteroid families.

\textit{The dedicated PTF spin-rate study} \citep[][; hereafter PTF-Chang]{Chang2014, Chang2015}. There
are 1,751 reliable rotation periods obtained from three dedicated surveys. Each survey used the
Palomar 48-inch Oschin Schmidt Telescope to observe 87 deg$^2$ in Mould-$R$ band with 20 min candence in the ecliptic plane. They were carried out in February 15--18 of 2013 and in January 6--9 and February 20--23 of 2014, respectively. The collected asteroid lightcurves were analyzed by the traditional second-order Fourier series method to find rotation period \citep{Harris1989}. The surveys have relatively lower ability to detect rotation period $> 12$ hr (see Figure 5 in \citet{Chang2015}). The data set mainly contains main-belt asteroids of hundred meter to 20 km in size.

We discard the rotation periods having difference $> 5$ \% between both data sets and obtain 173 unique asteroids in total for the Koronis family (see Table~\ref{table1}).

\section{Results and Discussion}
\begin{figure}
  \plotone{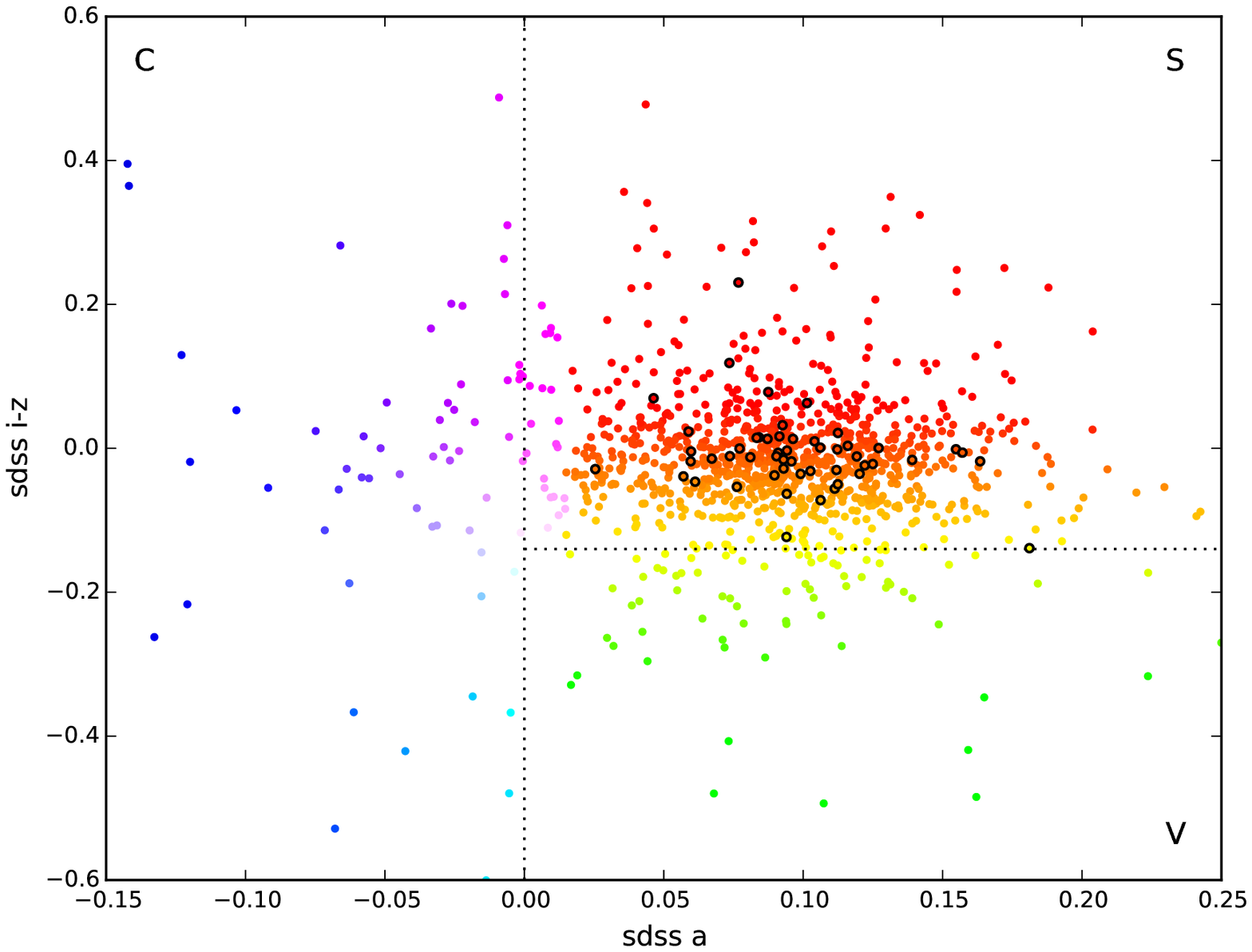}
  \caption{The spectral types of the Koronis family members obtained from SDSS. The colors indicate different spectral types. Black circles denote objects with their rotation periods determined from the PTF data.}
  \label{fig_sdss}
  \end{figure}

\begin{figure}
  \plotone{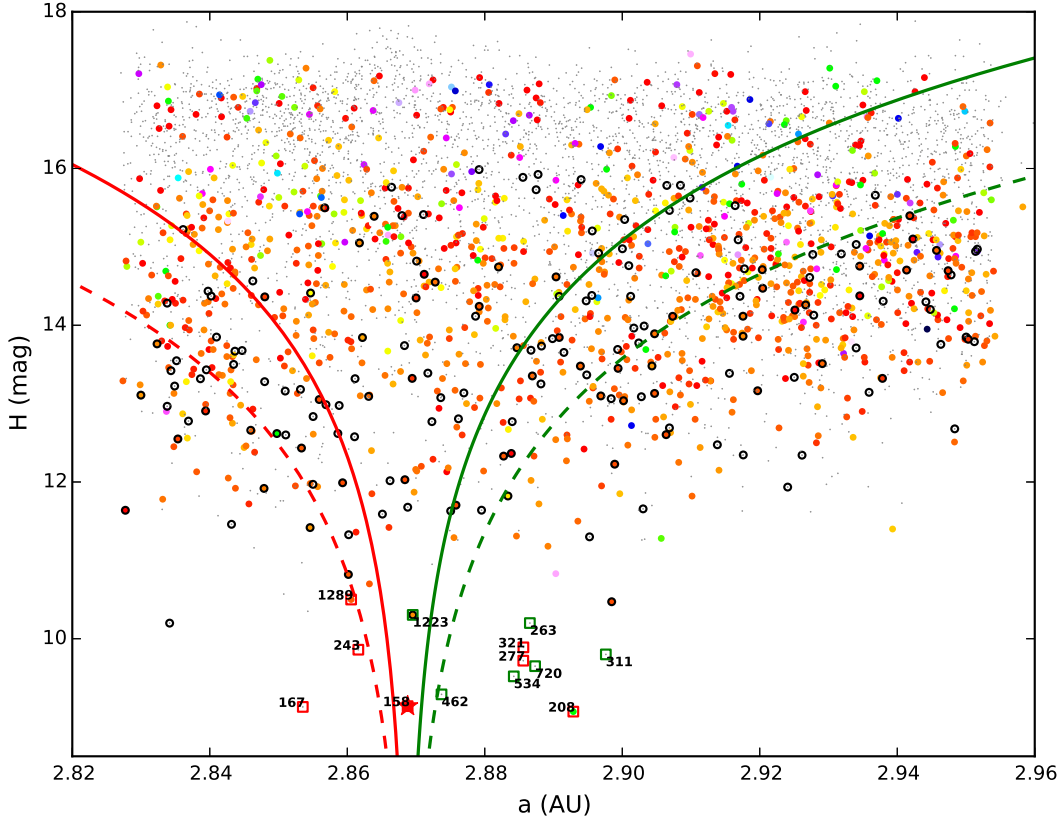}
  \caption{The plot of $a$ vs. $H$ of the Koronis family. The small gray dots are the members without SDSS colors and the objects with black circle are the members with available PTF rotation periods. The red star, (158) Koronis, is the center of the family. The solid and dashed curves indicate $\Delta a$ of $V_0 =$ 50 and 100 m/s, respectively, where green means the outer side of the family and red is the inner side. The members locating beyond $\Delta a$ might have been drifted in semimajor axis as a result of the Yarkovsky effect, where prograde and retrograde rotators would increase and decrease in semimajor axis, respectively. The green squares are prograde and the red squares are retrograde rotators from \citet{Slivan2002} and \citet{Slivan2009} with their numbers marked on the plot.}
  \label{fig_ah}
  \end{figure}

Among 5146 HCM selected Koronis family members, 1131 objects have SDSS colors. Their colors show a preferential concentration of S-type in Figure~\ref{fig_sdss}. This is expected for the Koronis family as pointed out by \citet{Cellion2002}.
Therefore, those non-S-type objects in the HCM selected Koronis family members are considered as interlopers, which only make up a very minor population (i.e., $<$ 15\%) and would have limited affect in the following analysis. The typical ``V'' shape of asteroid families is apparent in the plot of $a$ vs. $H$ of the Koronis family (see Figure~\ref{fig_ah}). The boundaries at $\sim2.83$ and $\sim2.95$ AU are due to the powerful 5/2 and 7/3 mean-motion resonances with Jupiter. We also notice that a large part of the sample of \citet{Slivan2002} are at least $\sim$2 mag brighter than ours. In Figure~\ref{fig_ah}, $\Delta a$ of $V_0 =$ 50 and 100 m/s are drawn in solid and dashed lines, respectively, where red represents the inner side of the Koronis family and green is the outer side. The family members locating at the outside the red and green lines are inner- and outer-side members, respectively.

\begin{figure}
  \plotone{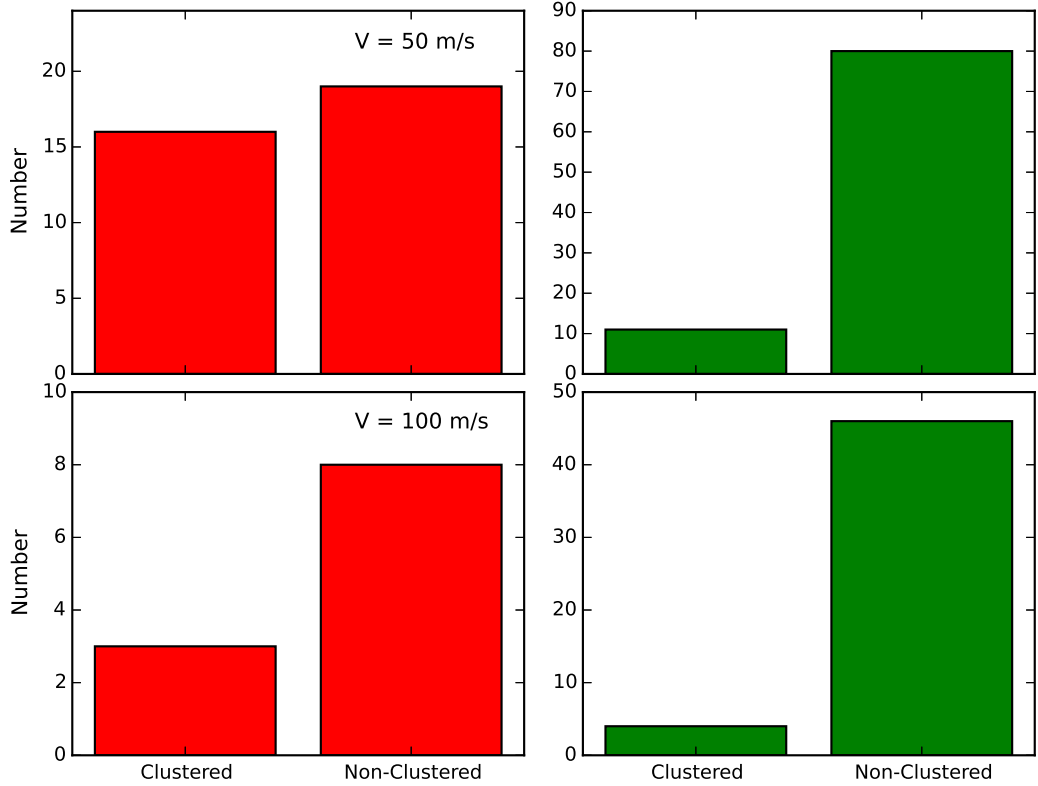}
  \caption{The bimodal distributions of clustered (i.e., the Slivan presents) and non-clustered rotation period of the Koronis members for $\Delta a$ of $V_0 =$ 50 (upper) and 100 m/s (lower). Green represents outer-side members and red is inner-side members. The clustered rotation period for outer-side members is between 7--10 hr and that for the inner-side members is $< 5$ and $> 13$ hr.}
  \label{fig_cluster}
  \end{figure}

To inspect rotation period clustering of the Slivan state, we use a bimodal distribution to divide the rotation periods into ``clustered'' (i.e., presenting the Slivan state) and ``non-clustered'' (i.e., not presenting the Slivan state). Therefore, the clustered rotation period for the inner-side members is $< 5$ or $> 13$ hr, and that for the outer-side members is between 7--10 hr. If the rotation period does not fulfill the aforementioned criteria, it is classified as non-clustered rotation period. The results are given in Figure~\ref{fig_cluster}. The ratios of clustered to non-clustered outer-side members are 0.14 for $V_0 = 50$ m/s and 0.09 for $V_0 = 100$ m/s. This indicates that most of outer-side members are probably not in the Slivan state. The ratios of clustered to non-clustered inner-side members are 0.84 for $V_0 = 50$ m/s and 0.38 for $V_0 = 100$ m/s. Although these ratios of inner-side members are higher than the outer-side members, its distributions are still dominated by non-clustered rotation periods. Therefore, we conclude that rotation period clustering is not observed in our results.

If the inner-/outer-side members are dominated by retrograde/prograde rotators as the general scenario of
Yarkovsky drifting described in section~\ref{method}, our method should work properly and consequently
rotation period clustering should be seen in our analysis on the Koronis family. This means that
determination of the sense of rotation in our sample would be helpful to verify whether
retrograde/prograde rotators are dominant in the inner-/outer-side members of the Koronis family.

\begin{figure}
  \epsscale{0.6}
  \plotone{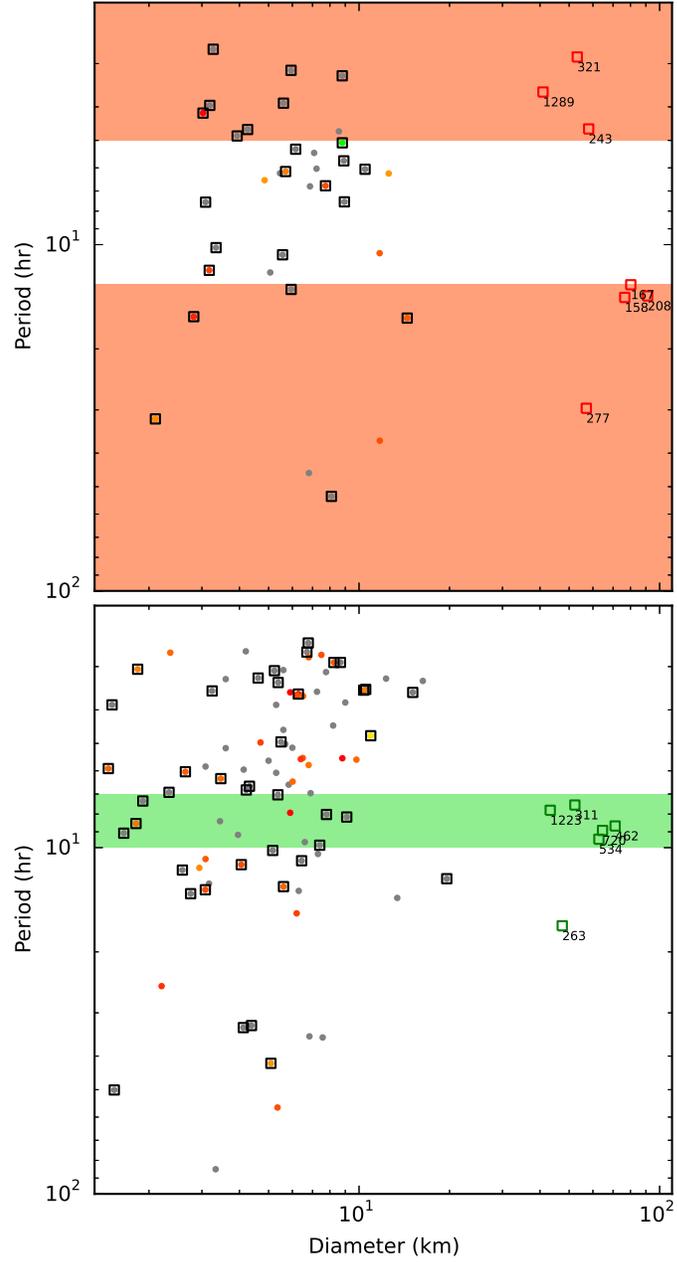}
  \caption{The plot of $D$ vs. rotation period of the Koronis family. The symbols and the colors are the same with Figure~\ref{fig_ah}. The red shadow in the upper panel indicates the rotation period clustering region for the inner-side members and the green shadow in the lower panel is that for the outer-side members. The black squares are the members between $\Delta a$ of $V_0 =$ 50 and 100 m/s.}
  \label{fig_dp}
  \end{figure}

We note that, the old Koronis family could have undergone a certain degree of collision after the initial
catastrophic break-up event, such that the spin states of the members could have altered and
consequently the dispersion in semimajor axis would not evolve in the way as described in Section~\ref{method}. Such intermittent collisional process capable of changing the rotational configuration is expected to be more frequent for the small members of the Koronis family. Figure~\ref{fig_dp} is the plot of $D$ vs rotation period and it shows that we have much smaller diameter in our sample than \citet{Slivan2002}. This might be the reason why the rotation period clustering is not observed in our result. Therefore, numerical simulation of Yarkovsky drifting for the Koronis family members including random collisional effect would be helpful to understand whether the scenario of the Slivan state should be applicable to the general population of for the Koronis family.

Since the Koronis family is relatively old, a population of interlopers in our sample could
jeopardize our method to separate prograde and retrograde rotators. However, the Koronis family
members with available SDSS colors display a concentration of S-type in a C-type dominant environment
(i.e., the outer main belt). Therefore, we believe that the population of interlopers should be minor
in the HCM selected members, unless these interlopers happen to be S-type asteroids.

\citet{Slivan2009} reported a ``stray'' Koronis family member, (263) Dresda. This object is a prograde rotator with $H = 10.4$ mag locating at the outer side. However, its spin status is not in the Slivan state (i.e., rotation period of 16.81 hr and obliquity of $16^\circ$). Because of the initial spin state and location of Dresda, the YORP effect
could not drive it into the Slivan state \citep{Slivan2009}. Such non-clustered spin state evolution
was also studied by \citet{Vokrouhlicky2003}. Therefore, the ``stray'' Dresda may provide another
scenario for spin state evolution for the Koronis family. Moreover, the YORP effect could also
destroy the Slivan state, especially it works more efficiently on small asteroid. We suspect that the
non-clustering rotation period distribution is probably caused by this mechanism.


\section{Summary and Conclusion}
Asteroid families are believed to
have been produced by catastrophic break-up events, forming swarms of fragments distributed in intervals of orbital proper elements $\Delta a$, $\Delta e$, $\Delta i$, reminiscent of the orbital elements of their parent bodies. In the subsequent orbital evolution, prograde and retrograde rotators would have drifted out of the $\Delta a$ to the far and near sides of the family by Yarkovsky effect, respectively. If the Slivan state exists in the bulk of Koronis family, the outer-side members would show rotation period clustering within 7-10 hr and the inner-side members would avoid rotation period interval of 5--13 hr \citep{Slivan2002}. With the large data set of rotation periods reported in PTF-Waszczak and PTF-Chang, we do not see rotation period clustering in the inner-/outer-side members of the Koronis family with $V_0 = $ 50 and 100 m/s. Available SDSS colors for the Koronis family members shows 85\% of them are of S-type, we therefore believe that our method cannot be seriously affected by the limited contribution of interlopers. It would be very important to measure the pole orientations of our sample, and thereby to verify our conclusion. Moreover, the ``stray'' spin state of the family member, (263) Dresda, perhaps represents another pathway for the spin state evolution of the Koronis
family. Finally it is noted that intermittent collisional process in the asteroid belt could lead to
further disruption and change in the spin states of small objects not accounted for by the secular
YORP effect.

\acknowledgments We thank the referee, Alberto Cellino, for his useful suggestions and comments to improve the content of the paper. We also thank Chung-Ming Ko for his English correction. This work is supported in part by the National Science Council of Taiwan under the grants MOST 104-2112-M-008-014-MY3 and MOST 101-2119-M-008-007-MY3, and also by Macau Science and Technology Fund No. 017/2014/A1 of MSAR.

\begin{deluxetable}{rrrrrrrrrrr}
\tabletypesize{\scriptsize} \setlength{\tabcolsep}{0.02in} \tablecaption{The Koronis family members
with available rotation periods obtained from PTF data sets. \label{table1}} \tablewidth{0pt}
\tablehead{ \colhead{Obj Number} & \colhead{$a$ (AU)} & \colhead{$e$} & \colhead{$sin(i)$} &
\colhead{$H$ (mag)} & \colhead{D (km)} & \colhead{SDSS a$^*$} & \colhead{SDSS i} & \colhead{Type} &
\colhead{Period (hr)} & \colhead{$\triangle m$ (mag)}} \startdata
     975 &  2.872 &  0.045 &  0.036 &  10.20 &   22.2$^a$ &      - &      - &   - &   7.22 &   0.20 \\
    1100 &  2.868 &  0.047 &  0.035 &  10.47 &   16.9$^a$ &      - &      - &   - &  14.53 &   0.10 \\
    1223 &  2.875 &  0.048 &  0.037 &  10.30 &   22.8$^a$ &      - &      - &   - &   7.83 &   0.28 \\
    1423 &  2.870 &  0.044 &  0.037 &  10.82 &   19.6$^a$ &   0.10 &  -0.05 &   s &  12.29 &   0.96 \\
    1497 &  2.918 &  0.080 &  0.035 &  11.30 &   16.3$^a$ &      - &      - &   - &   3.30 &   0.20 \\
    1618 &  2.870 &  0.043 &  0.036 &  11.68 &   17.5$^a$ &   0.09 &   0.02 &   s &  43.32 &   0.65 \\
    1802 &  2.872 &  0.044 &  0.039 &  11.46 &   12.4$^a$ &      - &      - &   - &   3.16 &   0.37 \\
    1912 &  2.860 &  0.045 &  0.037 &  11.66 &   26.3     &   0.14 &  -0.01 &   s &   4.63 &   0.47 \\
    1913 &  2.928 &  0.071 &  0.039 &  11.64 &   13.4$^a$ &      - &      - &   - &  13.97 &   0.27 \\
    1955 &  2.920 &  0.073 &  0.037 &  11.97 &    9.8$^a$ &   0.14 &  -0.02 &   s &   5.57 &   0.39 \\
    2123 &  2.868 &  0.048 &  0.036 &  11.33 &   14.5$^a$ &   0.09 &   0.00 &   s &  16.30 &   0.19 \\
    2498 &  2.865 &  0.046 &  0.034 &  12.35 &   11.3$^a$ &      - &      - &   - &   3.06 &   0.15 \\
    2620 &  2.880 &  0.047 &  0.036 &  12.62 &   16.9     &      - &      - &   - &   3.39 &   0.56 \\
    2713 &  2.883 &  0.046 &  0.038 &  11.42 &   15.1$^a$ &   0.09 &  -0.12 &   s &   3.56 &   0.61 \\
    2726 &  2.876 &  0.049 &  0.037 &  11.99 &   10.9$^a$ &   0.13 &   0.00 &   s &   4.75 &   0.26 \\
    2811 &  2.936 &  0.070 &  0.035 &  11.59 &   12.3$^a$ &      - &      - &   - &   3.25 &   0.32 \\
    2814 &  2.876 &  0.044 &  0.034 &  12.03 &    9.9$^a$ &      - &      - &   - &   3.37 &   0.18 \\
    2833 &  2.856 &  0.046 &  0.037 &  11.70 &   25.8     &   0.12 &   0.00 &   s &   3.49 &   0.11 \\
    2931 &  2.855 &  0.045 &  0.037 &  11.63 &   11.7$^a$ &   0.12 &  -0.06 &   s &  36.82 &   0.23 \\
    2953 &  2.857 &  0.043 &  0.038 &  11.64 &   12.6$^a$ &      - &      - &   - &   6.23 &   0.15 \\
    2985 &  2.866 &  0.046 &  0.033 &  11.92 &   10.5$^a$ &      - &      - &   - &   6.06 &   0.48 \\
    3131 &  2.859 &  0.049 &  0.037 &  11.94 &   23.2     &   0.15 &   0.00 &   s &  19.75 &   0.70 \\
    3377 &  2.882 &  0.043 &  0.037 &  12.47 &    7.7$^a$ &   0.12 &  -0.04 &   s &   6.76 &   0.33 \\
    3726 &  2.853 &  0.044 &  0.037 &  12.02 &   10.4$^a$ &   0.07 &  -0.01 &   s &   3.51 &   0.11 \\
    3791 &  2.884 &  0.047 &  0.037 &  11.82 &   11.7$^a$ &      - &      - &   - &  10.58 &   0.18 \\
    3975 &  2.883 &  0.049 &  0.037 &  12.23 &    9.1$^a$ &   0.09 &  -0.01 &   s &   8.16 &   0.71 \\
    4351 &  2.888 &  0.047 &  0.036 &  12.43 &    8.2$^a$ &      - &      - &   - &   2.92 &   0.27 \\
    4365 &  2.879 &  0.071 &  0.035 &  12.60 &    7.4$^a$ &      - &      - &   - &   9.84 &   0.69 \\
    4506 &  2.925 &  0.073 &  0.037 &  12.33 &    8.8$^a$ &   0.05 &   0.07 &   s &   5.52 &   0.76 \\
    4664 &  2.866 &  0.050 &  0.036 &  12.36 &   19.0     &      - &      - &   - &   3.11 &   0.31 \\
    4965 &  2.855 &  0.047 &  0.037 &  13.86 &    9.6     &      - &      - &   - &  12.55 &   0.35 \\
    5187 &  2.859 &  0.041 &  0.036 &  12.34 &    8.9$^a$ &      - &      - &   - &   7.52 &   0.30 \\
    5494 &  2.851 &  0.046 &  0.037 &  12.62 &    8.9$^a$ &      - &      - &   - &   5.73 &   0.98 \\
    5523 &  2.877 &  0.042 &  0.038 &  13.42 &    8.1$^a$ &      - &      - &   - &  53.33 &   0.31 \\
    5582 &  2.850 &  0.045 &  0.037 &  13.37 &    7.6$^a$ &   0.32 &  -0.35 &   v &   4.50 &   0.23 \\
    5677 &  2.869 &  0.042 &  0.040 &  13.22 &    8.8$^a$ &   0.08 &   0.02 &   s &   5.08 &   0.20 \\
    5759 &  2.874 &  0.050 &  0.036 &  13.68 &    7.0$^a$ &      - &      - &   - &   4.20 &   0.61 \\
    5941 &  2.891 &  0.044 &  0.036 &  13.25 &   12.6     &      - &      - &   - &  10.19 &   0.94 \\
    6006 &  2.891 &  0.044 &  0.037 &  12.78 &    7.5$^a$ &   0.11 &   0.00 &   s &   2.77 &   0.28 \\
    6210 &  2.887 &  0.043 &  0.035 &  12.58 &    8.7$^a$ &      - &      - &   - &   2.92 &   0.25 \\
    6262 &  2.888 &  0.043 &  0.037 &  12.60 &    7.8$^a$ &      - &      - &   - &   8.02 &   0.49 \\
    6686 &  2.848 &  0.045 &  0.036 &  12.68 &    8.6$^a$ &      - &      - &   - &   4.71 &   0.42 \\
    6829 &  2.860 &  0.040 &  0.036 &  12.98 &    7.3$^a$ &      - &      - &   - &  27.82 &   0.28 \\
    6980 &  2.846 &  0.043 &  0.037 &  12.97 &    8.8$^a$ &      - &      - &   - &   3.25 &   0.40 \\
    7028 &  2.932 &  0.070 &  0.035 &  12.77 &    7.3$^a$ &      - &      - &   - &  10.43 &   0.24 \\
    7120 &  2.879 &  0.041 &  0.040 &  13.09 &   13.6     &      - &      - &   - &   2.87 &   0.18 \\
    7210 &  2.864 &  0.052 &  0.036 &  12.55 &    8.2$^a$ &   0.09 &  -0.04 &   s &   6.24 &   0.29 \\
    7244 &  2.857 &  0.052 &  0.036 &  13.08 &    7.5$^a$ &      - &      - &   - &  35.03 &   0.83 \\
    7340 &  2.851 &  0.051 &  0.036 &  13.76 &    5.6$^a$ &      - &      - &   - &  10.70 &   0.76 \\
    7372 &  2.888 &  0.049 &  0.038 &  12.66 &   16.6     &      - &      - &   - &   2.56 &   0.19 \\
    8102 &  2.890 &  0.051 &  0.037 &  13.55 &    5.4$^a$ &      - &      - &   - &   3.33 &   0.32 \\
    8234 &  2.898 &  0.045 &  0.035 &  13.74 &   10.1     &      - &      - &   - &  33.09 &   0.49 \\
    8361 &  2.884 &  0.040 &  0.040 &  13.11 &    6.3$^a$ &   0.10 &   0.06 &   s &   3.60 &   0.75 \\
    8571 &  2.843 &  0.044 &  0.036 &  13.18 &    6.2$^a$ &      - &      - &   - &   5.30 &   0.54 \\
    8657 &  2.845 &  0.048 &  0.037 &  13.32 &    5.9$^a$ &      - &      - &   - &  13.47 &   0.67 \\
    9078 &  2.843 &  0.042 &  0.039 &  12.99 &    6.8$^a$ &      - &      - &   - &  45.65 &   0.44 \\
    9122 &  2.879 &  0.038 &  0.037 &  14.12 &    8.5     &   0.09 &  -0.01 &   s &   7.92 &   0.27 \\
    9306 &  2.861 &  0.053 &  0.036 &  12.81 &    7.2$^a$ &      - &      - &   - &   3.79 &   0.18 \\
    9337 &  2.846 &  0.050 &  0.037 &  13.32 &    5.9$^a$ &      - &      - &   - &   3.14 &   0.13 \\
    9726 &  2.848 &  0.039 &  0.037 &  14.82 &    6.1     &   0.10 &   0.01 &   s &   5.55 &   0.28 \\
    9916 &  2.947 &  0.069 &  0.038 &  13.16 &    6.2$^a$ &   0.10 &   0.01 &   s &  15.48 &   0.45 \\
    9999 &  2.874 &  0.053 &  0.037 &  12.91 &    7.1$^a$ &      - &      - &   - &   3.48 &   0.15 \\
   10025 &  2.951 &  0.077 &  0.036 &  12.69 &    8.2$^a$ &      - &      - &   - &   4.44 &   0.48 \\
   10052 &  2.870 &  0.037 &  0.038 &  13.09 &    6.1$^a$ &      - &      - &   - &   4.07 &   0.24 \\
   11762 &  2.840 &  0.042 &  0.036 &  13.13 &   13.4     &      - &      - &   - &   6.22 &   0.53 \\
   12186 &  2.898 &  0.047 &  0.037 &  13.77 &    9.9     &   0.12 &   0.00 &   s &  11.20 &   0.29 \\
   12356 &  2.897 &  0.050 &  0.035 &  13.14 &    6.0$^a$ &   0.10 &  -0.02 &   s &   6.45 &   0.38 \\
   12540 &  2.899 &  0.050 &  0.036 &  13.06 &    7.2$^a$ &      - &      - &   - &   3.55 &   0.67 \\
   12564 &  2.894 &  0.041 &  0.037 &  14.28 &    5.4$^a$ &      - &      - &   - &   7.04 &   0.44 \\
   12838 &  2.887 &  0.039 &  0.039 &  12.77 &   15.8     &      - &      - &   - &  10.91 &   0.48 \\
   13266 &  2.837 &  0.045 &  0.038 &  13.84 &    9.6     &      - &      - &   - &   4.85 &   0.64 \\
   13306 &  2.839 &  0.048 &  0.037 &  13.16 &    7.1$^a$ &      - &      - &   - &   5.43 &   0.41 \\
   13806 &  2.861 &  0.054 &  0.036 &  14.38 &    7.5     &      - &      - &   - &  20.70 &   0.26 \\
   14164 &  2.835 &  0.042 &  0.036 &  14.31 &    7.8     &   0.09 &   0.01 &   s &  11.86 &   0.71 \\
   14243 &  2.835 &  0.043 &  0.038 &  14.19 &    8.2     &      - &      - &   - &  10.20 &   0.73 \\
   14246 &  2.905 &  0.045 &  0.036 &  13.14 &    6.8$^a$ &   0.10 &  -0.03 &   s &   5.77 &   0.70 \\
   14382 &  2.906 &  0.046 &  0.036 &  13.96 &    5.8$^a$ &      - &      - &   - &   6.58 &   0.29 \\
   14779 &  2.903 &  0.043 &  0.035 &  13.32 &    6.8$^a$ &      - &      - &   - &  35.07 &   0.64 \\
   14977 &  2.895 &  0.039 &  0.038 &  14.71 &    4.9$^a$ &      - &      - &   - &  44.03 &   0.71 \\
   14992 &  2.887 &  0.038 &  0.040 &  13.85 &    5.6$^a$ &   0.06 &  -0.02 &   s &  12.96 &   0.47 \\
   15195 &  2.945 &  0.078 &  0.038 &  14.76 &    6.3$^a$ &      - &      - &   - &  13.33 &   0.41 \\
   15339 &  2.927 &  0.082 &  0.036 &  13.51 &    5.0$^a$ &      - &      - &   - &   5.61 &   0.25 \\
   15409 &  2.896 &  0.051 &  0.033 &  13.76 &    6.7$^a$ &      - &      - &   - &   2.72 &   0.13 \\
   15414 &  2.950 &  0.086 &  0.036 &  13.79 &    4.2$^a$ &      - &      - &   - &   2.71 &   0.21 \\
   15421 &  2.834 &  0.044 &  0.038 &  14.37 &    7.6     &      - &      - &   - &   7.54 &   0.35 \\
   15460 &  2.859 &  0.036 &  0.036 &  14.11 &    5.4$^a$ &      - &      - &   - &   7.27 &   0.22 \\
   15564 &  2.951 &  0.081 &  0.036 &  13.71 &    5.3$^a$ &      - &      - &   - &   6.08 &   0.38 \\
   16131 &  2.835 &  0.048 &  0.036 &  14.24 &    8.0     &      - &      - &   - &   2.73 &   0.49 \\
   16199 &  2.834 &  0.048 &  0.038 &  14.30 &    7.8     &      - &      - &   - &   3.96 &   0.73 \\
   16326 &  2.900 &  0.073 &  0.035 &  13.50 &    5.6$^a$ &      - &      - &   - &   3.07 &   0.28 \\
   16483 &  2.834 &  0.041 &  0.039 &  13.48 &    5.1$^a$ &      - &      - &   - &  12.03 &   0.17 \\
   16541 &  2.848 &  0.040 &  0.041 &  13.04 &   14.0     &   0.09 &  -0.04 &   s &   6.15 &   0.33 \\
   16626 &  2.892 &  0.037 &  0.037 &  13.69 &   10.3     &      - &      - &   - &   6.81 &   0.37 \\
   16909 &  2.844 &  0.054 &  0.037 &  13.05 &    6.9$^a$ &      - &      - &   - &   6.79 &   0.38 \\
   16917 &  2.902 &  0.040 &  0.034 &  13.39 &   11.9     &      - &      - &   - & 321.55 &   0.69 \\
   17163 &  2.885 &  0.055 &  0.038 &  13.89 &    4.9$^a$ &   0.06 &  -0.04 &   s &   4.11 &   0.23 \\
   17605 &  2.894 &  0.037 &  0.038 &  14.12 &    8.5     &   0.09 &  -0.02 &   s &   6.32 &   0.58 \\
   17694 &  2.840 &  0.038 &  0.037 &  13.67 &   10.4     &      - &      - &   - &   4.65 &   0.33 \\
   17779 &  2.830 &  0.042 &  0.039 &  13.39 &   11.9     &   0.09 &  -0.06 &   s &   6.51 &   0.40 \\
   18930 &  2.899 &  0.040 &  0.040 &  14.37 &    7.5     &   0.14 &   0.00 &   s &  13.23 &   0.79 \\
   19083 &  2.862 &  0.054 &  0.041 &  13.35 &   12.1     &   0.08 &  -0.05 &   s &   3.05 &   0.25 \\
   20774 &  2.900 &  0.043 &  0.041 &  13.48 &    6.5$^a$ &   0.16 &  -0.02 &   s &   3.65 &   0.17 \\
   21027 &  2.907 &  0.045 &  0.040 &  14.26 &    7.9     &      - &      - &   - &   3.53 &   0.27 \\
   21566 &  2.901 &  0.040 &  0.034 &  13.84 &    5.2$^a$ &      - &      - &   - &   3.08 &   0.25 \\
   21580 &  2.906 &  0.041 &  0.035 &  13.10 &    6.2$^a$ &   0.05 &   0.00 &   s &   3.63 &   0.28 \\
   21993 &  2.914 &  0.048 &  0.037 &  14.37 &    7.6     &      - &      - &   - &   5.83 &   0.68 \\
   22497 &  2.917 &  0.045 &  0.036 &  13.73 &   10.1     &      - &      - &   - &   5.95 &   0.26 \\
   22899 &  2.910 &  0.047 &  0.039 &  13.68 &    5.7$^a$ &      - &      - &   - &   5.02 &   0.14 \\
   23056 &  2.896 &  0.056 &  0.035 &  13.85 &    4.3$^a$ &      - &      - &   - &   6.65 &   0.42 \\
   23543 &  2.834 &  0.047 &  0.033 &  12.84 &    7.2$^a$ &      - &      - &   - &   6.03 &   0.22 \\
   23873 &  2.863 &  0.035 &  0.038 &  13.71 &    5.7$^a$ &      - &      - &   - &  13.86 &   1.23 \\
   24874 &  2.897 &  0.035 &  0.040 &  14.70 &    4.6$^a$ &      - &      - &   - &   3.24 &   0.23 \\
   25112 &  2.862 &  0.053 &  0.042 &  14.74 &    6.4     &   0.11 &  -0.07 &   s &  10.13 &   0.35 \\
   25171 &  2.918 &  0.042 &  0.035 &  14.47 &    7.2     &   0.06 &  -0.05 &   s &  11.44 &   0.33 \\
   25241 &  2.916 &  0.043 &  0.034 &  13.32 &    6.0$^a$ &      - &      - &   - &   5.15 &   0.58 \\
   25285 &  2.916 &  0.041 &  0.034 &  13.82 &    7.6$^a$ &      - &      - &   - &  35.34 &   0.76 \\
   27502 &  2.948 &  0.081 &  0.037 &  14.55 &    6.9$^a$ &      - &      - &   - &   6.96 &   0.34 \\
   30305 &  2.901 &  0.055 &  0.036 &  14.44 &    4.4$^a$ &      - &      - &   - &  32.64 &   0.57 \\
   31050 &  2.904 &  0.055 &  0.036 &  14.36 &    5.1$^a$ &   0.11 &  -0.06 &   s &  41.97 &   0.42 \\
   32873 &  2.855 &  0.038 &  0.039 &  14.90 &    5.9     &      - &      - &   - &   4.64 &   0.54 \\
   34087 &  2.828 &  0.053 &  0.038 &  14.41 &    7.4     &   0.12 &   0.07 &   s &   4.17 &   0.53 \\
   34469 &  2.905 &  0.049 &  0.040 &  14.70 &    6.5     &   0.07 &  -0.01 &   s &   6.04 &   0.63 \\
   35513 &  2.917 &  0.054 &  0.034 &  14.62 &    6.7     &      - &      - &   - &  13.58 &   0.53 \\
   36036 &  2.895 &  0.057 &  0.037 &  14.94 &    5.8     &      - &      - &   - &   9.46 &   0.25 \\
   36056 &  2.868 &  0.060 &  0.035 &  13.34 &   12.2     &      - &      - &   - &  14.48 &   0.30 \\
   36281 &  2.839 &  0.057 &  0.038 &  14.56 &    6.9     &   0.09 &   0.03 &   s &  16.14 &   0.24 \\
   37346 &  2.924 &  0.076 &  0.033 &  14.38 &    3.6$^a$ &      - &      - &   - &   3.26 &   0.44 \\
   38272 &  2.840 &  0.060 &  0.037 &  13.43 &    5.6$^a$ &      - &      - &   - &   3.90 &   0.15 \\
   38963 &  2.929 &  0.047 &  0.038 &  14.96 &    5.8     &   0.12 &  -0.02 &   s &   2.73 &   0.43 \\
   38992 &  2.918 &  0.054 &  0.037 &  14.98 &    5.7     &      - &      - &   - &   6.92 &   0.61 \\
   41797 &  2.871 &  0.034 &  0.038 &  14.64 &    6.7     &   0.08 &   0.23 &   s &  10.14 &   0.39 \\
   43883 &  2.836 &  0.059 &  0.034 &  15.66 &    4.2     &      - &      - &   - &  28.59 &   0.32 \\
   45259 &  2.900 &  0.059 &  0.037 &  14.75 &    6.3     &      - &      - &   - &  11.61 &   0.79 \\
   45303 &  2.926 &  0.051 &  0.037 &  14.13 &    8.4     &      - &      - &   - &   8.39 &   0.25 \\
   45586 &  2.908 &  0.059 &  0.033 &  13.83 &    9.7     &      - &      - &   - &   9.19 &   0.31 \\
   45628 &  2.873 &  0.034 &  0.042 &  15.22 &    5.1     &   0.03 &  -0.03 &   s &   3.48 &   0.39 \\
   46583 &  2.927 &  0.056 &  0.036 &  15.53 &    4.4     &   0.09 &  -0.03 &   s &   8.52 &   0.27 \\
   47691 &  2.944 &  0.066 &  0.032 &  13.28 &    5.3$^a$ &      - &      - &   - &   3.87 &   0.54 \\
   50493 &  2.841 &  0.063 &  0.037 &  15.40 &    4.7     &      - &      - &   - &   4.64 &   0.40 \\
   50811 &  2.937 &  0.076 &  0.032 &  14.65 &    3.6$^a$ &      - &      - &   - &   5.16 &   0.42 \\
   51987 &  2.935 &  0.090 &  0.036 &  14.61 &    6.8$^a$ &   0.08 &  -0.01 &   s &   2.82 &   0.22 \\
   52179 &  2.942 &  0.073 &  0.036 &  15.39 &    4.7$^a$ &   0.08 &   0.01 &   s &   4.97 &   0.81 \\
   53298 &  2.920 &  0.076 &  0.038 &  13.45 &    6.5$^a$ &   0.16 &  -0.01 &   s &   5.51 &   0.43 \\
   54272 &  2.832 &  0.063 &  0.038 &  15.20 &    5.1     &   0.11 &  -0.05 &   s &  31.86 &   0.58 \\
   55004 &  2.946 &  0.084 &  0.036 &  13.99 &    9.0$^a$ &      - &      - &   - &   3.81 &   0.39 \\
   56558 &  2.935 &  0.081 &  0.034 &  14.92 &    5.9$^a$ &   0.07 &   0.12 &   s &   3.56 &   0.24 \\
   60552 &  2.902 &  0.034 &  0.036 &  15.62 &    4.2     &      - &      - &   - &  29.24 &   0.56 \\
   60679 &  2.950 &  0.087 &  0.033 &  14.72 &    6.4$^a$ &   0.06 &   0.02 &   s &   5.55 &   0.87 \\
   62298 &  2.925 &  0.057 &  0.040 &  14.20 &    8.2     &      - &      - &   - &  84.86 &   0.51 \\
   63660 &  2.871 &  0.064 &  0.038 &  13.65 &    4.7$^a$ &      - &      - &   - &  14.85 &   0.21 \\
   63795 &  2.934 &  0.056 &  0.039 &  15.92 &    3.7     &      - &      - &   - &   3.87 &   0.62 \\
   66742 &  2.938 &  0.059 &  0.037 &  14.31 &    7.8     &      - &      - &   - &  12.70 &   0.23 \\
   69934 &  2.938 &  0.058 &  0.037 &  15.10 &    5.4     &   0.11 &   0.02 &   s &  25.11 &   0.46 \\
   71049 &  2.941 &  0.053 &  0.036 &  14.35 &    5.4$^a$ &   0.08 &  -0.00 &   s &  56.30 &   0.46 \\
   72653 &  2.920 &  0.064 &  0.037 &  14.37 &    7.6     &   0.12 &  -0.01 &   s &  10.79 &   0.46 \\
   76069 &  2.928 &  0.063 &  0.039 &  15.73 &    4.0     &      - &      - &   - &   9.08 &   0.68 \\
   76142 &  2.903 &  0.064 &  0.037 &  15.35 &    4.8     &      - &      - &   - &   5.50 &   0.70 \\
   78596 &  2.853 &  0.066 &  0.037 &  15.40 &    4.7     &      - &      - &   - &  28.51 &   0.67 \\
   79988 &  2.911 &  0.064 &  0.037 &  15.41 &    4.7     &      - &      - &   - &   7.33 &   0.41 \\
   81205 &  2.907 &  0.076 &  0.038 &  15.05 &    5.5$^a$ &      - &      - &   - &   4.95 &   0.78 \\
   82730 &  2.899 &  0.062 &  0.035 &  15.78 &    3.9     &   0.11 &  -0.00 &   s &  17.33 &   0.56 \\
   83422 &  2.952 &  0.081 &  0.034 &  14.67 &    6.6$^a$ &      - &      - &   - &   9.64 &   0.55 \\
   97059 &  2.890 &  0.066 &  0.037 &  14.97 &    5.7     &   0.11 &  -0.03 &   s &   3.94 &   0.41 \\
  104188 &  2.948 &  0.053 &  0.037 &  15.89 &    3.8     &      - &      - &   - &  50.10 &   0.83 \\
  112888 &  2.895 &  0.030 &  0.037 &  15.46 &    4.6     &      - &      - &   - &  35.36 &   0.40 \\
  113049 &  2.886 &  0.066 &  0.036 &  15.09 &    5.4     &      - &      - &   - &   6.44 &   0.51 \\
  113515 &  2.934 &  0.090 &  0.037 &  15.03 &    5.6$^a$ &      - &      - &   - &   4.57 &   0.69 \\
  116503 &  2.855 &  0.065 &  0.035 &  15.86 &    3.8     &   0.18 &  -0.14 &   s &  11.82 &   0.68 \\
  132830 &  2.946 &  0.061 &  0.037 &  15.99 &    3.6     &   0.12 &  -0.02 &   s &   5.91 &   0.40 \\
  133528 &  2.918 &  0.067 &  0.037 &  15.50 &    4.5     &   0.10 &  -0.04 &   s &   3.05 &   0.35 \\
  135223 &  2.942 &  0.085 &  0.038 &  14.91 &    5.9$^a$ &   0.09 &   0.08 &   s &   7.93 &   0.78 \\
  198605 &  2.907 &  0.067 &  0.037 &  15.79 &    3.9     &   0.06 &   0.00 &   s & 160.77 &   0.38 \\
  268623 &  2.903 &  0.066 &  0.039 &  15.76 &    4.0     &      - &      - &   - &   7.01 &   0.68 \\
\enddata
\tablecomments{$a$, $e$, $i$ and $H$ are adopted from the Nesvorny HCM Asteroid Families V2.0.
Rotation period and lightcurve amplitude $\Delta m$ are obtained from PTF data sets.}
\end{deluxetable}

\end{document}